\documentclass[fleqn,usenatbib]{mnras}

\usepackage{newtxtext,newtxmath}

\usepackage[T1]{fontenc}
\usepackage{ae,aecompl}

\usepackage{graphicx}    
\usepackage{amsmath}    
\usepackage{amssymb}    
\usepackage{xcolor}
\usepackage[titletoc]{appendix}
\newcommand{\degree}{$^{\circ}$}

\title{Velocity centroid gradients for absorbing media}

\author[Gonz\'alez-Casanova,  Lazarian, \& Burkhart]{
Diego F. Gonz\'alez-Casanova$^{1}$,
A. Lazarian$^{1}$
and Blakesley Burkhart$^{2}$
\\
$^{1}$Astronomy Department, University of Wisconsin-Madison, 475 North Charter Street, Madison, WI 53706-1582, USA\\
$^{2}$Center for Computational Astrophysics, Flatiron Institute, 162 Fifth Avenue, New York, NY 10010, USA\\
}

\date{Accepted 2018 October 9. Received 2018 October 9; in original form 2017 April 3}

\pubyear{2017}

\begin{document}
\label{firstpage}
\pagerange{\pageref{firstpage}--\pageref{lastpage}}
\maketitle

\begin{abstract} 
We explore how the velocity gradient technique (VGT) can be applied to absorbing media in the case of $^{13}$CO 2-1 emission. The VGT is a new way to trace magnetic fields in the plane of the sky using only spectroscopic observations. We apply the VGT to magnetohydrodynamic turbulence simulations that have been post-processed to include $^{13}$CO 2-1 emission and we calculate the velocity centroid gradients. We find that the velocity centroid gradients trace the projected magnetic field in media with different $^{13}$CO abundances, densities and optical depths. Our study opens up the possibility of using velocity centroid gradients to trace magnetic fields in molecular clouds using $^{13}$CO emission.
\end{abstract}

\section{Introduction}

Magnetic fields play a fundamental role in astrophysical environments, where they are important for both the diffuse and dense molecular components of the interstellar medium \citep[ISM; see][]{draine2011}. Magnetohydrodynamic (MHD) processes are significant contributors to the dynamics of giant molecular clouds (GMCs), as the magnetic field can regulate star formation \citep[]{crutcher2012, burkhart2015,mocz2017,2017A&A...607A...2S}. The importance of turbulence and magnetic fields for star formation is demonstrated by observational measurements of linewidth size relations, power spectra and Zeeman measurements \citep{ostriker2003, Heyer2004,mckee2007, Ballesteros-Paredes11a,crutcher2012}. Zeeman measurements also demonstrate that magnetic pressure can provide support against self-gravity, and that the magnetic field energy can be comparable to turbulent kinetic energy  \citep{crutcher2010, draine2011}. Given that GMC turbulence parameters such as the sonic and Alfv{\'e}nic Mach numbers ($M_S$ and $M_A$, respectively) affect the initial properties and outcomes of star-forming regions, it is vital to characterize the medium \citep{burkhart2013}.

The velocity gradient technique (VGT) provides a new way of tracing magnetic fields. The VGT was first introduced by \citet[][henceforth GL17]{casanova2017} and quantified and extended by \citet{yuen2017,lazarian2018a} and \citet{lazarian2018b}. Recently, \citet{yuen2018} have presented a detailed comparison with other statistical techniques for measuring magnetic fields in the ISM. \citet{hu2018} have improved the precision of tracing magnetic fields in the VGT using principal component analysis. In addition, \cite{lazarian2018a} have shown that the VGT is able to measure the distribution of magnetization given Alv\'en Mach numbers $M_A$, while \citet{yuen2018b} have demonstrated a way to trace the distribution of sonic Mach number $M_s$ with H{\sc i} data. These recent works suggest that the VGT is a very promising tool for studying magnetic fields and turbulence.

However, in all of the aforementioned numerical studies, line absorption effects were disregarded. Similarly, the applications were mostly focused on H{\sc i} data for which optical depth effects and absorption can be disregarded.\footnote{The only exception is the application of the incarnations of the VGT technique, namely, Velocity Channel Gradients (VChGs) to 13CO Vela C data in \citep{lazarian2018a}.} Therefore, it is essential to understand the effects of absorption for application of the VGT to molecular line observations.

The foundations of the VGT are rooted in MHD turbulence theory \citep[see][for a review and ref. therein]{brandenburg2013} as well as the theory of the statistics of position-position-velocity (PPV) space that is available in observations \citep{lazarian2000}. The modern picture of MHD turbulence scaling was first presented by \citet{goldreich1995}, henceforth GS95, who suggested the theory of strong MHD turbulence. The alignment of the eddy motions with the local direction of the magnetic field - an essential point for the VGT technique - was predicted by \citet{lazarian1999}. In a turbulent magnetized fluid, velocity fluctuations trace magnetic field fluctuations surrounding the turbulent eddies, which can be inferred from velocity gradients. MHD theory predicts that, in the presence of turbulence, magnetic fields reconnect over one eddy turnover time \citep{lazarian1999}. Therefore, there is no resistance to motions that mix magnetic field lines in the perpendicular direction. At the same time, there is magnetic tension, which resists the bending motions of magnetic fields, so most of the energy is channelled into eddies that mix magnetic fields with minimal bending. The amplitude of the turbulent velocity gradients increases with decreasing spatial scales. As a result, the gradient measurement traces magnetic fields at the smallest scale resolved by a telescope. We note that the concept of eddy alignment with respect to the magnetic field has been convincingly demonstrated numerically \citep{cho2000, maron2001, cho2002a}. The results of the GL17 study demonstrated the ability of velocity centroid gradients (VCGs) to trace magnetic fields and to determine magnetic field strength using an approach similar to the well-known technique of \citet{chandrasekhar1953}, hereafter the C-F method.

The success of VCGs as a new way to study the ISM motivates us to revisit the problem and to address the issue of whether the ability of VCGs to trace magnetic fields in diffuse media will be significantly affected by self-absorption. The effects of self-absorption can become important for H{\sc i}, and they are definitely important for CO, which we investigate in this paper. In this work, we study the application of the VGT to $^{13}$CO 2-1 emission. We limit ourselves to the case where the gradients of velocities arising from self-gravity are small compared with the gradients induced by MHD turbulence. We expect the former to be dominant near the centres of self-gravitating regions (e.g. in the vicinity of forming stars), while for vast expanses of diffuse molecular gas we expect the MHD scaling to be applicable \citep{yuen2017b,burkhart2015,2018ApJ...856..136P,2017ApJ...834L...1B,2018arXiv180105428B}.

The study of the VGT presented here is done with different MHD conditions (compressibility and magnetization) and cloud properties ($H_2$ density and $^{13}$CO abundance). We change the density and $^{13}$CO abundance because different molecular clouds have different den- sities and abundances.

Our paper is organized as follows. In Section \ref{chap:VGT-CO-sec:theory}, we explore the theoretical approach of the VGT in PPV data. 
In Section \ref{chap:VGT-CO-sec:num}, we describe the numerical code and set-up of the simulations.
In Section \ref{chap:VGT-CO-sec:results}, we present the VGT in the presence of molecular emission. 
In Section \ref{chap:VGT-CO-sec:comp}, we compare with other techniques, and we discuss our results and conclusions in Section \ref{chap:VGT-CO-sec:conclusion}.

\section{Theoretical Considerations} \label{chap:VGT-CO-sec:theory}

In this section, we outline the basic MHD theory that is essential to the understanding of the VGT. MHD turbulence can have so-called weak and strong regimes \citep{brandenburg2013}, which refer to the strength of non-linear eddy interactions. The strength of interactions increases as the cascade proceeds to smaller scales (i.e. the strong non-linear regime). Sub-Alfv\'enic turbulence initially demonstrates a weak non-linear regime. However, provided that the inertial range of turbulence (i.e. the range from the injection scale to the dissipation scale) is sufficiently large, sub-Alfv\'enicc turbulence transfers to the strong non-linear regime at small scales \citep[see][]{lazarian1999}.

In the strong turbulence regime, GS95 Alfv\'enic turbulence is characterized by eddy-like turbulent motions. The turbulent motion leads to eddies being elongated along the magnetic field that surrounds the eddy \citep{lazarian1999, cho2000, maron2001}. The magnetic field therefore presents a natural frame of reference to measure the scales of the eddies, with the perpendicular $l_\perp$ and parallel $l_\parallel$ dimensions of the eddies defined relative to the magnetic field. These physical scales are related by the `critical balance relation'
\begin{equation}
l^{-1}_\parallel V_A \approx  l^{-1}_\perp u_l    \;,
\end{equation}
where $V_A$ is the Alfv{\'e}n speed and $u_l$ is the eddy velocity. With the turbulence injection scale and velocity ($L$ and $V_L$, respectively), we can write for sub-Alfv{\'e}nic turbulence (i.e. the turbulence with the Alfv{\'e}n Mach number $M_A=V_L/V_A<1$ \citep[see][]{lazarian1999}:
\begin{equation}
l_\parallel \approx L \Big(\frac{l_\perp}{L}\Big)^{2/3} M_A^{-4/3} \;.
\label{chap:VGT-CO-eq:eddy-velocity}
\end{equation}

This shows that the disparity between $l_{\bot}$ and $l_{\|}$ is increasing with decreasing scale. Furthermore, this scale dependence is also present in the velocity field:
\begin{equation}
u_l \approx V_L \Big(\frac{l_\perp}{L}\Big)^{1/3} M_A^{1/3} \;.
\end{equation}

This implies that the velocity gradients increase with decreasing scale, which is what the VGT is intended to capture. Therefore, in the context of GS95 turbulence, it is natural to expect the largest velocity gradients to be perpendicular to the local direction of the magnetic field. In this way, velocity gradients trace magnetic fields in MHD turbulence.

Unfortunately, three-dimensional (3D) velocity gradients cannot be obtained directly from observations. Velocity centroids must instead be constructed from spectral line data (i.e. PPV data), which trace a turbulent velocity field along the line of sight. In this paper, we study the properties of VCGs in the presence of line absorption. A similar study of other measures, such as velocity channel gradients \citep[see][]{lazarian2018a}, will be provided elsewhere.

The velocity centroid $C$ is the first-order moment of the intensity map. In the case of radio observations, the velocity centroid is measured by the antenna temperature ($T$):
\begin{equation}
C(\mathbf{X}) \propto \int v_z T(\mathbf{X},v_z) dv_z\;,
\end{equation}
where the integration is performed over the line of sight (LOS) component of velocity ($v_z$) and $\mathbf{X}$ is the position in the plane of sky (POS). By themselves, velocity centroids are a well-known measurement and their use for tracing magnetic fields has been previously demonstrated. For instance, \citet{esquivel2005} and \citet{burkhart2014} suggested that the anisotropies of the velocity centroid correlation functions could be used to study the magnetic field direction and \citet{casanova2018} showed the effects of the magnetic field using the Tsallis statistics on the velocity centroid. Velocity and intensity centroids, using the VGT, have been used to detect star-forming regions in the presence of self-gravity \citep{yuen2017b}.

\section{Simulated Observations} \label{chap:VGT-CO-sec:num}

We use the data cubes from \citet{burkhart2013}. These data cubes were obtained using 3D numerical simulations generated by the compressible MHD code presented by \citet{cho2002}, which were post-processed to account for opacity effects using the \texttt{SimLine-3D} software package  \citep{ossenkopf2002}. The details of the codes are described in the aforementioned papers as well as in \citet{burkhart2013b} and \citet{correia2014}. It is known that different turbulence drivers in simulations can lead to different results \citep{2010A&A...512A..81F,yoon2016}. In our case, we use a solenoidal driver, at wave-scale $k$ equal to about 2.5 (i.e. 2.5 times smaller than the size of the box). With this approach, we try to minimize the forcing of density structures.

The synthetic observational data that we analyse assume a box size of $\sim$5~pc (corresponding to a cell size of $\sim$0.0097~pc), and a temperature of 10 K. The assumed telescope beam that produces the synthetic PPV intensity maps is 18 arcmin FWHM, with the synthetic observation made at a distance of 450 pc with a velocity resolution of 0.05~km~s$^{-1}$.  The number density of $H_2$, $n_{H_2}$, and the $^{13}$CO abundance, $x_{co}$, (the number-density ratio of $^{13}$CO/H$_2$), are changed by factors of 30 from the normal parameters (with five cases for each of the five MHD models; see Table \ref{chap:VGT-CO-table:data}). The normal parameters are set at a density of $n_{H_2}$ = 275~cm$^{-3}$ and abundance $x_{co}$ = 1.5$\times10^{-6}$, representing typical values for GMCs. These changes allow investigation of line-saturation effects in the VGT.

We describe our simulation parameter set-up in Table \ref{chap:VGT-CO-table:data}. Columns 2-4 gives the sonic and Alfv\'en Mach numbers and the gas pressure of the MHD simulations ($M_S$, $M_A$ and $p$). Columns 5-9 give the LOS averaged optical depth ($\tau$). Column 5 corresponds to the normal case $n_{H_2}$ = 275~cm$^{-3}$ and $^{13}$CO abundance, $x_{co}$ = 1.5$\times10^{-6}$.  
Column 6 corresponds to a lower density case $n^1_{H_2}$ = $n_{H_2}/30$.  Column 7 corresponds to a high density case, $n^2_{H_2}$ = $30\cdot n_{H_2}$.  
Column 8 corresponds to a low $^{13}$CO abundance, $x^3_{co}$ = $x_{co}/30$.  Column 9 corresponds to a high $^{13}$CO abundance, $x^4_{co}$ = $30\cdot x_{co}$.

The $M_S$ reported in Table \ref{chap:VGT-CO-table:data} comes from MHD simulations where 3D information is available. Because spectroscopic observations cannot measure the 3D velocity dispersion and present opacity broadening (especially for the high optical depth cases), the measured Ms estimations differ. The deviations between the two can be accounted for by a factor that depends on the density and magnetic field. In this case they change by a factor ranging from 0.8 to 1.17. For the exact values, see \citet{correia2014}, who measured the sonic Mach number from spectral line data under the influence of opacity broadening.

\begin{table}
\centering
\caption{ Parameters of the simulations used with their average optical depth}
\label{chap:VGT-CO-table:data}
\begin{tabular}{c|c|c|c|c|c|c|c|c}
Model &  &  &  & Case normal & Low density & High density & Low $x_{co}$ & High $x_{co}$\\
&     &      &    & $ n_{H_2}= 275$ \& $x_{co}$ = 1.5$\times10^{-6}$ & $n^1_{H_2}$ = 9 & $n^2_{H_2}$ = 8250 & $x^3_{co}$ = 5$\times10^{-8}$ & $x^4_{co}$ = 4.5$\times10^{-5}$\\
 & $M_A$ & $M_S$ & p & $\tau$ & $\tau$ & $\tau$ & $\tau$ & $\tau$ \\
 \hline\hline
A   & 0.7         & 0.7          &1    & 3.22          & 0.323        & 49.8    & 0.232      & 52.2             \\
B   & 0.7         & 1      &0.7    & 3.48          & 0.355       & 54.7      & 0.252     & 57.2              \\
C   & 0.7         & 2      &0.1    & 3.52          & 0.348       & 54.1      & 0.235     & 57.2              \\
D   & 0.7         & 4      &0.05    & 3.21          & 0.344       & 54.5      & 0.207     & 57.4              \\
E   & 0.7         & 7    &0.01    & 3.45         & 0.376        & 62.0     & 0.213     & 65.0                 \\
\end{tabular}
\bigskip\\
Column 1 indicates the letter for the model for the MHD simulation given by the parameters in Columns 2-4, with column 4 being the pressure. In Columns 5-9, we show the map-averaged optical depths, $\tau$, for the different cases of our radiative transfer parameter space applied to the MHD simulations. This produces a total of 5 MHD models with 5 cases each.  For our radiative transfer parameter space, we vary both the scaling of the number density of H$_2$ (in units of cm$^{-3}$) and the molecular abundance ($^{13}$CO/H$_2$, denoted with the symbol $x_{co}$) in order to change excitation and optical depth. To represent the typical parameters of a molecular cloud, we choose standard values for density, $n_{H_2}$ = 275~cm$^{-3}$ , and abundance, $x_{co}$ = 1.5$\times10^{-6}$ (what we will refer to as the normal case setup shown in Column 5). The cloud parameters of density and abundance are changed by a factor of 30 larger and smaller, giving the values of Columns 6-9.
\end{table}

\section{Alignment of velocity gradients and magnetic field} \label{chap:VGT-CO-sec:results}

We employ the method suggested and tested by \citet{yuen2017} to estimate the VCGs. The velocity centroid map $C$ is interpolated using the bicubic spline approximation over a rectangular mesh, adding 10 extra cells between each of the original points. At each of the original cells with position $\mathbf{X}$, the direction of VCG is given by $\mathbf{X'}$ by estimating $\nabla(\mathbf{X})$. $\mathbf{X}$ and $\mathbf{X'}$ are points in the POS from the data cubes. $\nabla(\mathbf{X})$ is estimated with points $\mathbf{X'}$ defined in an annulus with a radius of 10 cells around $\mathbf{X}$. $\nabla(\mathbf{X})$ is defined as
\begin{equation}
\nabla(\mathbf{X}) =  \rm{max}\bigg\{ \frac{|C(\mathbf{X})- C(\mathbf{X}+\mathbf{X'})|}{|\mathbf{X'}|} \bigg\} \;.
\end{equation}
This method of calculating VCGs provides better accuracy and a more numerically robust approach than the G17 approach.
We compare the magnetic field direction inferred from the VCGs to the actual direction (obtained directly from the simulations) of the magnetic field by projecting it on to the POS:
\begin{eqnarray}
\mathbf{B}_x(\mathbf{X})  \propto  \int \mathbf{B}_x(\mathbf{X},z)   dz \;, \hfill \nonumber \\
\mathbf{B}_y(\mathbf{X})  \propto  \int \mathbf{B}_y(\mathbf{X},z)   dz  \;, \hfill \nonumber\\
\label{chap:VGT-CO-eq:mag-proj}
\end{eqnarray}

Here, $z$ is measured along the LOS and the mean magnetic field is perpendicular to the LOS. In the case where the mean magnetic field is not perpendicular to the LOS, any measurement of the alignment decreases. Future work should analyse the effects of the LOS orientation in more detail.

The VGT provides angle measurements (i.e. inferred directions of the magnetic field). In order to obtain the statistical properties of these angular measurements, we need to use circular statistics. Linear statistics will give an ambiguous or incorrect result due to the $2\pi$ periodicity intrinsic to angular data. To measure the spreads of the angle distributions, our primary tool will be the circular standard deviation, $S$.

The circular standard deviation is defined as:
\begin{equation}
\label{chap:VGT-CO-eq:sig}
S = \frac{1}{2}\Big\{ -2ln(R)\Big\} ^{1/2} \;,
\end{equation}
with $R$, the mean resultant length defined as:
\begin{equation}
\label{chap:VGT-CO-eq:R}
R = \frac{1}{n}\Bigg\{ \Big(\sum\limits_{i=1}^n cos(2\theta_i) \Big) ^2 + \Big(\sum\limits_{i=1}^n sin(2\theta_i) \Big)^2 \Bigg\} ^{1/2} \;,
\end{equation}
where $\theta_i$ is the angle of each data point and $n$ is the number of data points. The angles have to be multiplied by 2 and the spread divided by 2, given that angular measurements with the VGT are obtained from pseudo-vectors rather than actual vectors. In other words, the angle periodicity is $\pi$ and therefore the equations have to be modified to account for the different periodicity. $R$ describes how clustered the data is and relates to the variance of the data. Based on Eq.\ref{chap:VGT-CO-eq:R}, for $R\in$[0,1], $R=1$ in perfectly clustered data and $R=0$ in a uniform distribution. Following that, for $S\in[0,\infty)$, $S=0$ for clustered data and $S=\infty$ for a uniform distribution \citep{mardia1999}.

\begin{figure*}
\centering\includegraphics[width=0.8\linewidth,clip=true]{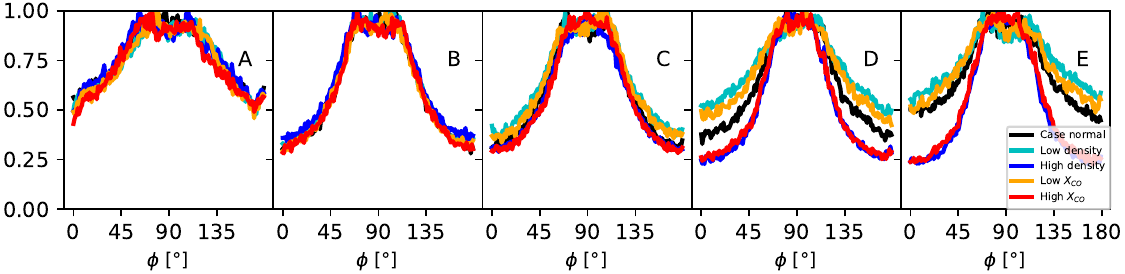}
\caption{Histogram of the angle distribution between the velocity centroid gradient and the magnetic field for all models. Each panel corresponds to one model and has the corresponding letter (see Table \ref{chap:VGT-CO-table:data}). $M_S$ increases from left to right. The different colors correspond to the different optical depths (variations in H$_2$ density and $x_{co}$).  The histograms are set to have 1 as the maximum value to compare with different sample sizes.  The VCG maps are smoothed using a Gaussian kernel with a standard deviation of 2 to attenuate any outliers}
\label{chap:VGT-CO-fig:hist-uc}
\end{figure*}

Fig. \ref{chap:VGT-CO-fig:hist-uc} shows the histograms of the angle distributions between VCGs and $\mathbf{B}$ given by the five different MHD initial conditions and the five cloud radiative transfer conditions. Each panel corresponds to a MHD model and the different colours correspond to the different $n$ and $x_{co}$ values. The histograms measuring the angle distributions are made after a Gaussian smoothing kernel with a standard deviation of two cells (0.0194 pc) is applied to the VCG data.

\begin{figure}
\centering\includegraphics[width=0.8\linewidth,clip=true]{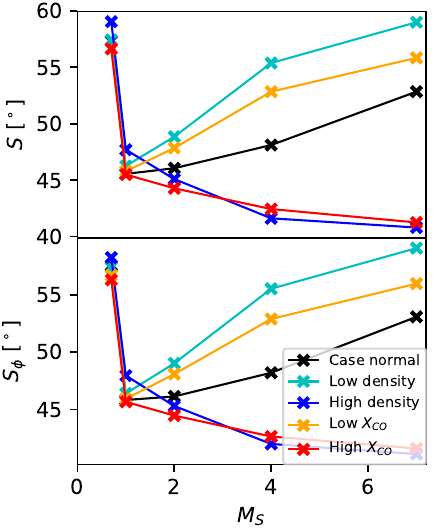}
\caption{ {\it Top Panel}: The circular standard deviation of the VCG, $S$ as a function of $M_S$. {\it Bottom Panel}: The circular standard deviation of the angle between the VCG and the magnetic field, $S_\phi$ as a function of $M_S$. The different colours correspond to the different cases (variations in density and $x_{co}$). {\it Black} is the normal conditions, {\it cyan} is the low density case (case 1), {\it blue} is the high density case (case 2), {\it orange} is the low $x_{co}$ (case 3), and {\it red} is the high $x_{co}$ (case 4).}
\label{chap:VGT-CO-fig:sig-ms}
\end{figure}

To quantify the alignment between the projected magnetic field and the VCG, we measure the angle between the two vectors, $\phi$, and the spread of the angle distribution, $S_\phi$, using the circular standard deviation (Eq. \ref{chap:VGT-CO-eq:sig}). Figure \ref{chap:VGT-CO-fig:sig-ms} shows $S_\phi$ as a function of the sonic Mach number.

Fig. \ref{chap:VGT-CO-fig:sig-ms} presents two different trends for subsonic and supersonic motions. The different trends are caused because the underlying physics has changed. When the medium is supersonic, it has shocks. Shocks produce a gradient in velocity at the shock front. This velocity gradient is different from the one produced by eddy motions, and therefore does not trace the magnetic field. Because of this, the overall alignment between the velocity gradients and the magnetic field decreases and is reflected in the spread of the angle distribution measured by $S$ and $S_\phi$.

The measurement of $\phi$ (and therefore of $S_\phi$) requires knowledge of the magnetic field. However, we can independently measure the spread of an angle distribution relative to its mean - in this case, the direction of the VCG. The corresponding measurement $S$ is the standard deviation of the VCG distribution and is independent of the magnetic field. $S$ and $S_\phi$ are different quantities that measure the width of the distribution, both with ($S_\phi$) and without ($S$). $S$ is the only quantity available from observations, where there is no information about the magnetic field.

\begin{figure}
\centering\includegraphics[width=0.8\linewidth,clip=true]{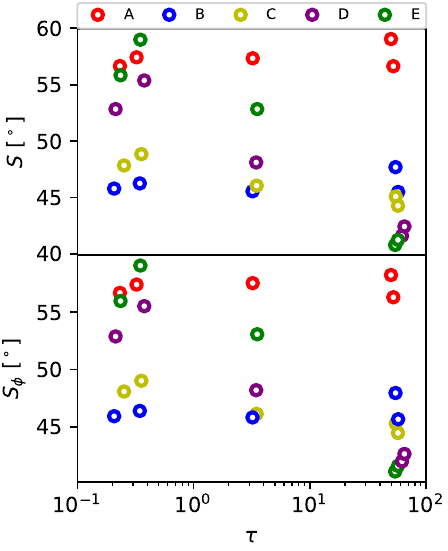}
\caption{ {\it Top Panel}: The circular standard deviation of the VCG, $S$ as a function of the optical depth, $\tau$. {\it Bottom Panel}: The circular standard deviation of the angle between the VCG and the magnetic field, $S_\phi$ as a function of $\tau$. The different colours correspond to the different models, labeled by their letter (see Table \ref{chap:VGT-CO-table:data}).}
\label{chap:VGT-CO-fig:sig-tau}
\end{figure}

Fig. \ref{chap:VGT-CO-fig:sig-tau} shows the dependence of $S$ and $S_\phi$ on the optical depth $\tau$. The values for the optical depth are shown in Table \ref{chap:VGT-CO-table:data} and correspond to the different cloud environments. It is clear that at higher optical depth (for supersonic models), the alignment is better as $S$ becomes smaller. This happens because the optical depth reduces the amount of the medium observed, which relates to the number of eddies in the LOS. The fewer the eddies, the fewer the con- volutions needed when performing the projection into the plane of the sky, and hence there is a better alignment. Furthermore, in the trans-sonic and supersonic turbulence cases the circular standard deviation is roughly 45\degree. Therefore, any measurements of the magnetic field with the VGT will be independent of the sonic Mach number.

\subsection{Gaussian Fit} \label{chap:VGT-CO-subsec:gauss}

The angle distribution for all cases and models shows a peak at $\sim$90\degree , implying that most cells have a good correspondence between the VCG and the magnetic field (see Fig. \ref{chap:VGT-CO-fig:hist-uc}). However, the spread in the distribution implies a natural dispersion due to statistical processes and the nature of MHD media \citep{esquivel2010}. In order to reduce the impact of the distribution’s natural dispersion and to understand better the local-global properties, we divided the data into sub-blocks.

\begin{figure*}
\centering\includegraphics[width=0.8\linewidth,clip=true]{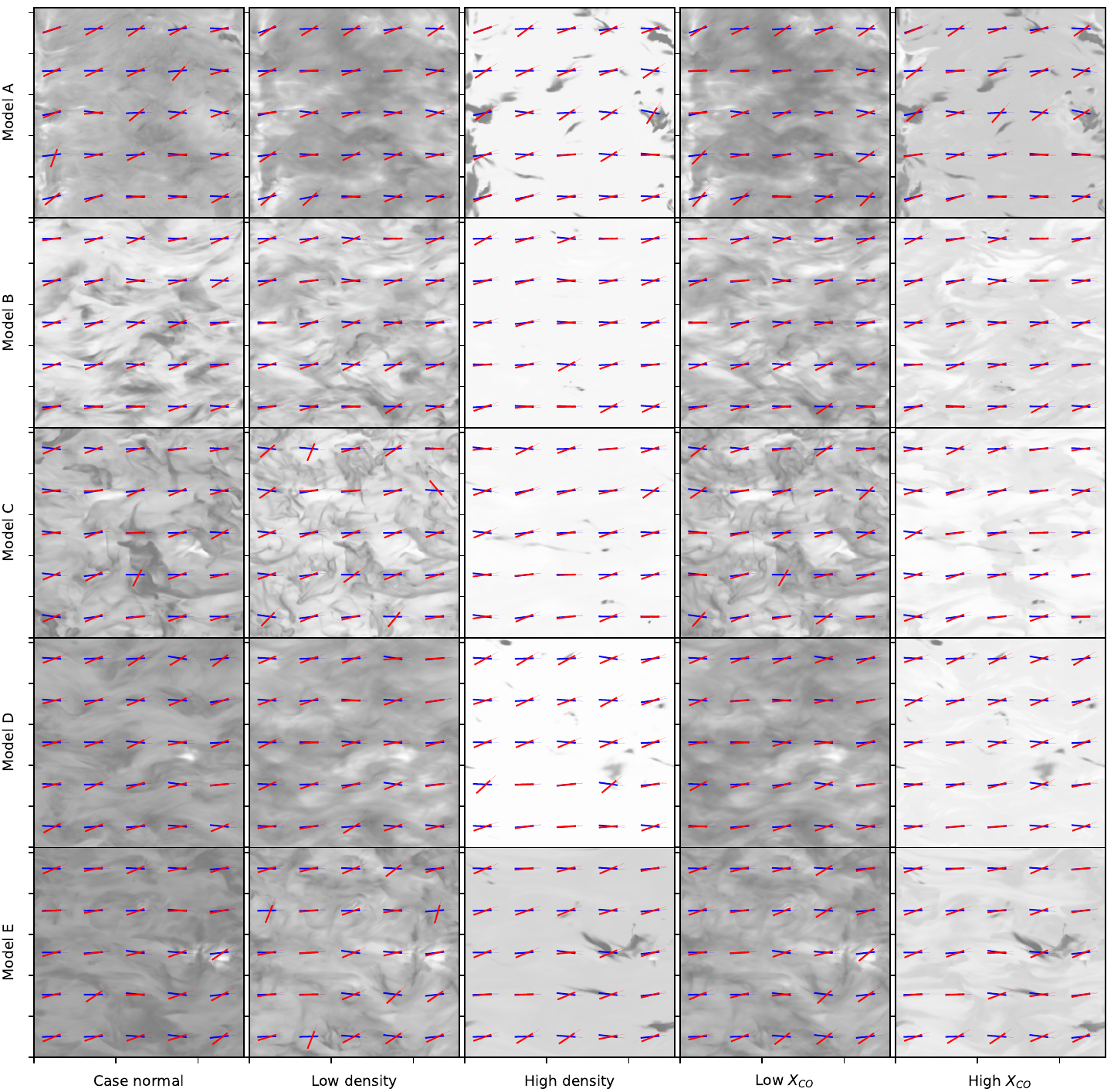}
\caption{The intensity map (log-scaled) for each of the different models and cases. In {\it red} the vector field of the VCG rotated 90\degree to match the magnetic field direction and in {\it blue} the direction of the POS magnetic field.  The pseudo-vectors plotted correspond to a Gaussian fit (subsection \ref{chap:VGT-CO-subsec:gauss} with block sizes of 100$^2$ cells.  Their respective values for the AM are seen in Fig. \ref{chap:VGT-CO-fig:reductionfactor}. Each row corresponds to a different model, and each column to a case.}
\label{chap:VGT-CO-fig:gauss-fitt}
\end{figure*}

At each sub-block, we found the peak of the distribution for both the VCG and the magnetic field. Then, for each sub-block, we assigned the most common direction (i.e. the peak) as the only direction for each of the measurements in that sub-block. The direction obtained represents a local-global property (and therefore the direction fluctuates between sub-blocks). The intensity map (in log space) and the local-global directions are found in Fig. \ref{chap:VGT-CO-fig:gauss-fitt}. This process to obtain the local-global direction of the magnetic fields makes the VGT a statistical technique, in the sense that it does not give the actual direction of the magnetic field at each cell, but still gives a local picture. With this process, the alignment between the VCG and the magnetic field increases because it neglects the width of the distribution and only uses its peak.

The size of each sub-block is dependent on the local alignment. As seen in \citet{yuen2017}, depending on the conditions of the sub-block, more or fewer cells are needed to obtain a Gaussian-like distribution to determine the maximum. If there are too few cells, a distribution will not arise from the data and hence the assignment of where the most common direction will be is ill-defined. However, once there are a sufficient number of points to determine the shape of the distribution, the addition of extra points will not gain any new information and will only lose spatial information. In this case, it is complicated to determine where these limits lie, as they depend on the compressibility and magnetization that produces the alignment. Fig. \ref{chap:VGT-CO-fig:gauss-fitt} presents this analysis with a sub-block size of 100$^2$ cells.

To assess the effects of the sub-block size in the calculations we estimated the Alignment Measure (AM). The AM is analogous to the Rayleigh factor used in dust alignment theory and has been used in the context of the VGT \citep{greenberg1968}:

\begin{equation}
AM = 2\Big\langle cos(\phi)^2-\frac{1}{2} \Big\rangle \;,
\end{equation}
 where $\phi$ is the angle between the VCG and the magnetic field, and $AM$ ranges from -1 to 1, with $AM$ = 1 for a parallel/antiparallel configuration and $AM$ = -1 for a perpendicular configuration. Like directional statistics, the alignment measure accounts for the circular properties of the data.
 
We first determine five different block sizes to subdivide the simulations. The smallest block size is 16$^2$ cells and presents enough cells to be confident that in every sub-block the peak of the distribution is properly defined. The largest block size is a quarter the size of the simulation box to keep spatial variability, a key property of the VGT.

\begin{figure}
\centering\includegraphics[width=0.8\linewidth,clip=true]{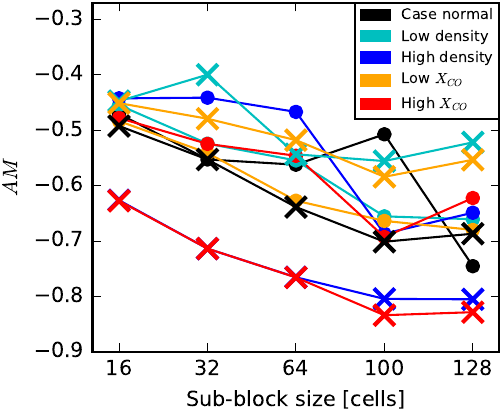}
\caption{The Alignment Measure, as a function of the sub-block size for models A ({\it crosses}) \& E ({\it dots}). The different colours correspond to the different cases (variations in density and $x_{co}$). {\it Black} is the normal conditions, {\it cyan} is the low density case (case 1), {\it blue} is the high density case (case 2), {\it orange} is the low $x_{co}$ (case 3), and {\it red} is the high $x_{co}$ (case 4).  The AM for the other models follows the behavior of models A and E with their values in between them.}
\label{chap:VGT-CO-fig:reductionfactor}
\end{figure}

After dividing the simulations by the different block sizes and obtaining the peak (angle) of the distribution at each of them, we estimated the alignment measure. Fig. \ref{chap:VGT-CO-fig:reductionfactor} shows the alignment measure for two models (A and E) and each of their five cases. The rest of the simulations show a similar trend and their values are in between the limiting cases of simulations A and E. It is clear that for the sub-Alfv\'enic regime, AM decreases, implying a better perpendicular alignment as the sub-block size increases until there is little gain at the largest block sizes. For all the sub-block sizes, a good alignment ($AM \ll0$) is achieved. Overall, the larger the sub-block, the better the alignment, but even at small sizes there is a good correspondence.

\subsection{Analog of the Chandrasekhar-Fermi Technique for the VGT} \label{chap:VGT-CO-subsec:c-f}

The C-F method is a technique to determine the intensity of the magnetic field in the ISM using two approximations: MHD motions are assumed to be Alfv\'enic, and the dispersion of the magnetic field is traced by the dispersion in the polarization measurements. Subsequent studies have analysed the method to better constrain the polarization dispersion and the empirical coefficient of the method \citep{ostriker2001, falceta2008, houde2009, novak2009}:
\begin{equation}
B_{CF} = C_{CF} \sqrt{4 \pi n} \frac{\sigma_v}{\sigma_\phi}  \;,
\label{chap:VGT-CO-eq:cf}
\end{equation}
where $C_{CF}$ is the empirical coefficient of proportionality, $\sigma_v$ is the dispersion in the velocity measured by the velocity centroid, $\sigma_\phi$ is the dispersion of the dust polarization, $n$ is the density and $B_{CF}$ is the estimated strength of the magnetic field.

G17 expanded the C-F method, incorporating the dispersion in the velocity gradient instead of the dispersion of the dust alignment. The physics of the approach remains, but now the dispersion of the magnetic field is measured by the VCG rather than by dust polarization. In the classical C-F method, $C_{CF} \sim 0.5$, while for the modified method for the VGT $C_{CF,VGT} \sim 1.29$ (with no self-absorption). We now expand it to self-absorption media, in particular $^{13}$CO emission with $C_{CF,VGT}\sim 0.1$. We use the circular standard deviation $S$ when estimating the spread of the angle dispersion.

To obtain $C_{CF,VGT}$, we solved Equation \ref{chap:VGT-CO-eq:cf} for $C_{CF,VGT}$ in the form:

\begin{equation}
C_{CF,VGT} = \frac{|B|}{\sqrt{4 \pi n}}  \frac{S}{\sigma_v}  \;,
\label{chap:VGT-CO-eq:cf2}
\end{equation}
where $|B|$ is the magnitude of the POS magnetic field obtained from
the simulations, $S$ is the circular standard deviation of the velocity centroid, $\sigma_v$ is the the spread of the velocity centroid estimated without self-absorption directly form the MHD simulations and n is the density.

As seen in \citet{correia2014}, the velocity dispersion measurements are affected by opacity effects. Therefore, when measuring the velocity dispersion, it is important to understand the opacity of the medium. This affects not only our modified approach to the C-F method but also the classical method using dust polarization \citep{crutcher2012}.

\begin{figure}
\centering\includegraphics[width=0.8\linewidth,clip=true]{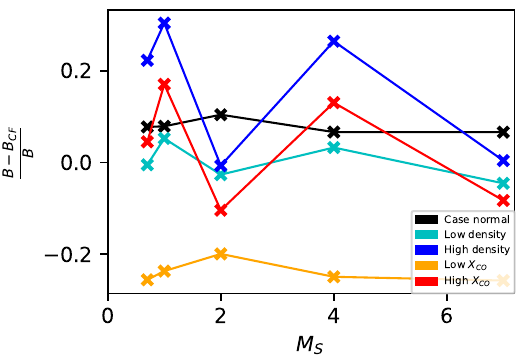}
\caption{Uncertainties in the estimates of the magnetic field strength using the modified C-F method as a function of $M_S$. The circular standard deviation was used to measure the width of the velocity gradient in the C-F method. The different colours correspond to the different cases (variations in density and $x_{co}$). {\it Black} is the normal conditions, {\it cyan} is the low density case, {\it blue} is the high density case, {\it orange} is the low $x_{co}$, and {\it red} is the high $x_{co}$.}
\label{chap:VGT-CO-fig:cf}
\end{figure}

Because opacity effects introduce uncertainty, we have chosen to use our MHD data cubes that neglect opacity (i.e. without radiative transfer). Thus, the values in Fig. \ref{chap:VGT-CO-fig:cf} reflect an analysis without opacity effects, measuring the velocity dispersion without self absorption using the C-F method. To see how the velocity dispersion is affected by the CO fraction and density for different sonic Mach numbers, see \citet{correia2014}. One way to approach this measurement is by obtaining the properties of the velocity field with linewidths that are not saturated (e.g. C$^{18}$O).

The densities presented in Table \ref{chap:VGT-CO-table:data} are post-processed using the radiative transfer code. These densities differ from the density derived from the MHD simulations used in the calculation of $C_{CF}$.

To obtain $C_{CF,VGT}$ , we determine the value for each of the different cases (as we have the information from the simulations). The value of $C_{CF,VGT}$ is then averaged over all of the values estimated individually for each model and case, as each model and case has the same magnetic field strength ($M_A\sim 0.7$). Fig. \ref{chap:VGT-CO-fig:cf} shows the relative errors of our C-F method using the VCG.

\section{Comparison with other techniques} \label{chap:VGT-CO-sec:comp}

Estimating the strength and direction of the magnetic field in GMCs is important for understanding both star formation and GMC dynamics. The direction of the magnetic field can be obtained by polarization measurements. The alignment of dust grains with the magnetic field that produces the polarization depends on the dust properties, making it dependent on the dust model. Furthermore, at high optical depths, there is no non-degenerate method for resolving the direction of the magnetic field, limiting the applicability of dust alignment theory \citep{lazarian2007, lazarian2015}.

For optically thin media, a comparison of the VGT and other statistical techniques, such as the correlation function anisotropy technique \citep{lazarian2002,esquivel2005,burkhart2014} or the principal component anisotropy analysis technique \citep{brunt2002,brunt2002a}, was provided by \citet{yuen2018}. They demonstrated the practical advantages of the VGT. Our present study, demonstrating that the VGT is capable of tracing magnetic fields in absorbing media, suggests that the technique is of superior utility for studies of molecular clouds using species where the effects of self-absorption are not negligible.

The importance of the VGT arises from its ability to measure the local direction of the magnetic field independent of dust polarization or more global measurements. The VGT produces both the strength and direction of the magnetic field using only spectroscopic data. As shown in this work, the VGT is efficient at tracing the magnetic field in synthetic observations with high optical depths, making it a very valuable tool to understand the physics of the GMCs.

\section{Conclusions} \label{chap:VGT-CO-sec:conclusion}

In this work, we extend the VGT to tracing magnetic fields in absorbing media. The VGT is based on the fact that eddies align with the local 3D magnetic field, creating eddies whose velocity gradients are perpendicular to the direction of the field. We find the following.

(i) The VGT is able to trace magnetic fields in sub-subAlfv{\'e}nicc, and subsonic and supersonic regimes for molecular emission with self-absorption.

(ii) Similar to H{\sc i}, a block average increases the accuracy of the magnetic field direction measurement in the presence of molecular emission from $^{13}$CO~2-1.

(iii) We extended the use of the C-F method to estimate the level of magnetization in molecular media. Unlike in the case of H{\sc i}, further studies are required to properly measure and fully understand the effects opacity in the measurement of linewidths.

\section*{Acknowledgements}

We acknowledge helpful exchanges with Ka Ho Yuen, with regards to the data analysis. We thank Zach Pace and Julie Davis for insightful discussions. Partial support for DFGC was provided by CONACyT (Mexico). AL was supported by the National Science Foun- dation AST grant 1816234 as a distinguished visitor PVE/CAPES appointment at the Physics Graduate Program of the Federal University of Rio Grande do Norte and thanks the INCT INEspa\c{c}o and Physics Graduate Program/UFRN, at Natal, for hospitality. BB is grateful for support from the Harvard ITC Postdoctoral Fellowship and the Simons Foundation Center for Computational Astrophysics.




\bibliographystyle{mnras}
\bibliography{biblio-grad2}{}




\bsp    
\label{lastpage}
\end{document}